\documentclass[letterpaper, 10 pt, conference]{ieeeconf}  

\IEEEoverridecommandlockouts                              
\usepackage[utf8]{inputenc}
\usepackage{amsthm}

\usepackage[noadjust]{cite}

\let\labelindent\relax
\usepackage{enumitem}

\usepackage{comment}

\usepackage{amsmath,amssymb, color}

\newcommand{\R}{\ensuremath{\mathbb{R}}}

\newcommand{\Rlo}{\ensuremath{\mathbb{R}_{\geq 0}}}

\newcommand{\Zo}{\ensuremath{\mathbb{Z}_{\geq 0}}}
\newcommand{\Zp}{\ensuremath{\mathbb{Z}_{> 0}}}
\newcommand{\Z}{\ensuremath{\mathbb{Z}}}

\definecolor{bleucit}{rgb}{0.2,0.4,0.6}

\definecolor{blue_cv}{rgb}{0.09,0.35,0.78}

\newcommand{\dom}{\ensuremath{\text{dom}\,}}

\newcommand{\sign}[1]{\ensuremath{\text{sign}{(#1)}}}

\newtheorem{thm}{\textnormal{\textbf{Theorem}}}

\newtheorem{rem}{\textnormal{\textbf{Remark}}}

\usepackage{tikz}
\usetikzlibrary{shapes,arrows,calc}
\definecolor{MyGreen}{RGB}{50,140,80}

\usepackage{ circuitikz }
\usetikzlibrary{arrows}

\title{\LARGE \bf
   Analysis of a Simple Neuromorphic Controller for Linear Systems: \\ A Hybrid Systems Perspective
}

\author{E. Petri \ \ \  K. J. A. Scheres \ \ \ E. Steur \ \ \ W. P. M. H. Heemels 
\thanks{Elena Petri, Koen Scheres, Erik Steur and Maurice Heemels are with the Department of Mechanical Engineering,	Eindhoven University of Technology, The Netherlands.
        {\tt\small \{e.petri, k.j.a.scheres, e.steur, m.heemels\}@tue.nl}}%
}

\begin{document}

\maketitle
\thispagestyle{empty}
\pagestyle{empty}

\begin{abstract}
In this paper we analyze a neuromorphic controller, inspired by the leaky integrate-and-fire neuronal model, in closed-loop with a single-input single-output linear time-invariant system. The controller  consists of two neuron-like variables and generates a spiking control input whenever one of these variables reaches a threshold. The control input is different from zero only at the spiking instants and, hence, between two spiking times the system evolves in open-loop. 
Exploiting the hybrid nature of the integrate-and-fire neuronal dynamics, we present a hybrid modeling framework to design and analyze this new controller. 
In the particular case of single-state linear time-invariant plants, we prove a practical stability property for the closed-loop system, we ensure the existence of a strictly positive dwell-time between spikes, and we relate these properties to the parameters in the neurons. The results are illustrated in a numerical example. 
%
\end{abstract}

\section{Introduction}

The advantages of brain-inspired technology, such as energy efficiency, robustness, accuracy,  and adaptability, have motivated many novel research topics in the last years. 
In particular, the pioneering work of Carver Mead \cite{mead1990neuromorphic}, highlights the inefficiency of digital computers compared to biological systems. This work kick-started the field of neuromorphic engineering, where novel technologies for computing, controlling, sensing and actuating systems, mimicking the structure and function of neurons and the (human) brain have been developed, see e.g., \cite{gallego2022event, van2018organic, krauhausen2021organic, shrestha2022survey}  and the references therein. 



%
%

The key features of neuron activity and communication are related to the generation and processing of spikes, where the information is contained in the presence or absence of a spike, 
rather than in its magnitude. 
Different models were presented in the literature to model the neuron dynamics that generate these spikes, see e.g., \cite{schuman2017survey,izhikevich2004model, izhikevich2007dynamical, drion2015neuronal} for surveys. In particular, neuron models can be divided in two categories based on the approach used to describe the spiking behavior. The first one consists in describing the neuron dynamics using only continuous-time dynamical systems, where spikes are captured by specific (smooth) nonlinearities, as, for instance, in the Hodgkin–Huxley \cite{hodgkin1952quantitative}, the Fitzhugh-Nagumo \cite{fitzhugh1961impulses} or the Hindmarsh-Rose \cite{hindmarsh1984model} neuron models. The second possible approach, which is the one we adopt in this paper, consists in describing the neuron dynamics by
continuous-time dynamical systems with discrete spikes, see, e.g., \cite{izhikevich2010hybrid}. 
%
In the latter approach, one of the simplest models to describe the neuron behavior is the (leaky) integrate-and-fire neuron model \cite{lapicque1907recherches,abbott1999lapicque, izhikevich2010hybrid}, \cite[Chapter 4]{gerstner2002spiking}, which consists in combining a first-order differential equation, describing the dynamics of the membrane potential of the neuron between spikes, and a mechanism that resets the membrane potential whenever a certain threshold is reached. Spikes are thus events that are triggered at precise moments in time and the continuous-time dynamics in between spikes determines the times when these events are triggered, which naturally leads to the possibility of using hybrid systems frameworks, e.g., \cite{heemels2010hybrid,goebel2012hybrid} to model neuron dynamics. 

In the field of control engineering, some of the first works illustrating the potential of brain-inspired control techniques are \cite{deweerth1990neuron, deweerth1991simple}, where the authors designed a proportional derivative neuromorphic controller for a DC motor. It is shown that the neuron servo controller works reliably when regulating the speed of the electric motor both at high and low reference speeds, unlike a conventional analog controller, which fails at low reference speeds because of the presence of friction. 
The interesting and promising results in \cite{deweerth1990neuron, deweerth1991simple} are unfortunately missing a formal technical analysis of the properties guaranteed by the controller.
More recently, a neuromorphic controller design has been proposed in \cite{ribar2021neuromorphic,sepulchre2022spiking}, where a positive feedback and a negative feedback are combined to obtain a mixed feedback controller that describes the mixed analog and digital nature of neurons. In particular, the positive feedback generates the spikes, while the negative feedback is used for regulation.
%
Other research provides analysis tools for controllers whose input comes from neuromorphic (event-based) sensors (such as the event-based camera), see, e.g., \cite{singh2016stabilization, singh2018regulation}. However, in this case, the control input is a continuous-time signal and not a spiking one and the controller is not designed by taking inspiration from the operation of the brain or from neuron dynamics. 

 
As previously mentioned, the generation of spikes 
can be perceived as the triggering of an event. Therefore, neuromorphic control has a strong connection with event-based control, for which many works have been published in recent years, see, e.g., \cite{aastrom1999comparison, tabuada2007event, heemels2012introduction, postoyan2014framework}. The main motivation of event-triggered control is to reduce the number of transmissions over a digital network compared to standard time-triggered periodic sampling, while still satisfying stability and performance properties. Thereto, the events are used to decide when a sample should be transmitted to obtain the desired stability and performance properties in an otherwise continuous control loop with continuously varying control signals taking any real value (not countable). 
In contrast, neurons, and consequently neuromorphic (spiky) controllers, operate through spikes, thereby controlling only through the time occurrences of events. The key information is therefore linked only to the triggering times.
Interestingly, the original inspiration for event-based control design comes from the event-based nature of biological systems and the control through events using pulses \cite{aastrom1999comparison, aaarzen1999simple}. 
%
%
In addition, control strategies where the control input has a saturated magnitude have been proposed in the literature, such as the quantized control in e.g., \cite{brockett2000quantized,liberzon2003hybrid, nesic2009unified}  and the ternary control in e.g., \cite{de2009robust}. In contrast to our work, these approaches do not consider spiking control actions and the controller design is not inspired by neurons. 
The focus of this work is to propose a formal framework to analyze spiky controllers, that are based on the leaky integrate-and-fire neuron model \cite{lapicque1907recherches,abbott1999lapicque, izhikevich2010hybrid}, \cite[Chapter 4]{gerstner2002spiking}, for linear time-invariant systems, which exploits the hybrid nature of neuron dynamics.
In particular, we consider a neuromorphic controller consisting of two neuron-like variables in closed-loop with a single-input single-output linear plant. 
As explained in \cite{izhikevich2010hybrid}, the leaky integrate-and-fire neuron model is only a threshold model and lacks the capability to generate spikes. To overcome this drawback and to better approximate the neuron behavior, in this work we add  
a mechanism that fires a spike whenever one of these variables reaches the threshold and is thus reset. The system is therefore controlled through spiking control inputs (such as the DC motor example mentioned above \cite{deweerth1990neuron, deweerth1991simple}) with fixed amplitude different from zero only at some discrete-time (triggering) instants.
%
%

The first main contribution of this work is the formulation of the structure of the neuron-inspired controller for a single-input single-output plant. We will exploit the hybrid nature of the neuronal dynamics and develop a suitable hybrid modeling setup for the closed-loop system. Specifically, we model the overall system as a hybrid system using the formalism of \cite{goebel2012hybrid}, where a jump corresponds to the occurrence of a spiking control action. 

Our second contribution is that, for the particular case of single-state linear time-invariant plants, we prove a practical stability property of the closed-loop system, which is not trivial in view of the structure of the proposed neuron-inspired controller and several technical difficulties have to be overcome. Indeed, the system evolves in open-loop except for some discrete instants and the control action has a fixed amplitude. Therefore, the system is controlled only through the time occurrence of events. As such, the design of the leaky integrate-and-fire neuron-like parameters, which essentially determine the times when spiking control actions occur, is 
 instrumental to guarantee a stability property of the system state, especially when the open-loop plant is not stable.
In addition, we prove the existence of a strictly positive dwell-time and thus we guarantee that the time between two consecutive spiking control actions is always greater than or equal to a strictly positive constant, thereby ruling out the Zeno phenomenon. 
Finally, we provide
	a qualitative relationship between the controller parameters (threshold, amplitude of the spiking control action and leaking parameter) and the guaranteed ultimate bound, region of attraction and dwell-time.
%
%
Compared to the existing works on neuromorphic control in the literature, this work proposes a novel modeling perspective and different tools to formally analyze a neuromorphic spiky controller with stability guarantees based on a simple hybrid model of neuronal dynamics. We illustrate these results also using a numerical case study. 
As the analysis is already technically challenging, we consider a simple class of systems, but our ambition for future work is to extend this framework to more general classes of systems, which may require the use of multiple pairs of neurons.

%
%


%

The remainder of the paper is organized as follows. Preliminaries are reported in Section~\ref{Notation}. The neuromorphic controller and the hybrid modeling framework are given in Section~\ref{ProblemStatement}, while Section~\ref{MainResult} presents the main analysis and design. A numerical example is given in Section~\ref{Example}. Finally, Section~\ref{Conclusions} concludes the paper and presents possible future research directions. The proof of the main result is given in the Appendix.

\section{Preliminaries}\label{Notation} 

The notation $\R$ stands for the set of real numbers, $\Rlo:= [0, +\infty)$ and $\R_{> 0}:= (0, +\infty)$. 
We use $\Z$ to denote the set of integers, $\Zo:= \{0,1,2,...\}$ and $\Zp:= \{1,2,...\}$. 
We consider hybrid systems in the formalism of \cite{goebel2012hybrid}, namely
\begin{equation}
\mathcal{H} \;:\; \left\{
\begin{array}{rcll}
\dot x &=& F(x), & \quad x \in \mathcal{C}, 
\\
x^+ & \in & G(x),  &\quad x \in \mathcal{D},
\end{array}
\right.
\label{eq:hybridSystemNotation}
\end{equation}
where 
$\mathcal{C}\subseteq \R^{n_x} $ is the flow set, 
$\mathcal{D}\subseteq \R^{n_x}$ is the jump set,
$F:\R^{n_x} \to \R^{n_x}$ is the flow map and $G:\R^{n_x} \to \R^{n_x}$ is the jump map. 
We consider hybrid time domains as defined in \cite[Definition 2.3]{goebel2012hybrid} and we use the notion of solution for system \eqref{eq:hybridSystemNotation} as advocated in \cite[Definition 2.6]{goebel2012hybrid}.
Given a solution $x$ for system \eqref{eq:hybridSystemNotation} and its hybrid time domain $\dom x$,  
$\sup_{j}\dom x := \sup \{j\in \Zo: \exists \, t\in \Rlo \textnormal{ such that } (t,j)\in \dom x
\}$.  

\section{Neuromorphic controller and hybrid modeling framework}\label{ProblemStatement}

	In this section we present the neuromorphic controller and we propose a new hybrid modeling framework for the closed-loop system. 

\begin{figure}
\begin{center}
\tikzstyle{blockB} = [draw, fill=blue!30, rectangle, 
minimum height=2em, minimum width=3em]  
\tikzstyle{blockG} = [draw, fill=MyGreen!40, rectangle, 
minimum height=2em, minimum width=3em]
\tikzstyle{blockR} = [draw, fill=red!40, rectangle, 
minimum height=2em, minimum width=3em]
\tikzstyle{blockO} = [draw,minimum height=1.5em, fill=orange!20, minimum width=4em]
\tikzstyle{input} = [coordinate]
\tikzstyle{blockW} = [draw,minimum height=1.5em, fill=white!20, minimum width=2.5em]
\tikzstyle{input1} = [coordinate]
\tikzstyle{blockCircle} = [draw, circle]
\tikzstyle{sum} = [draw, circle, minimum size=.3cm]
\tikzstyle{blockSensor} = [draw, fill=white!20, draw= blue!80, line width= 0.8mm, minimum height=10em, minimum width=13em]

\begin{tikzpicture}[auto, node distance=2cm,>=latex , scale=0.70,transform shape] 
	
	\node [input, name=stateFeedback] {};
	\node [input, right of= stateFeedback, node distance=1cm] (stateFeedbackRight){};
	\node [input, above of= stateFeedbackRight, node distance=1cm] (stateFeedbackUp){};
	\node [input, below of= stateFeedbackRight, node distance=1cm] (stateFeedbackDown){};
	
	\node [blockW, right of=stateFeedbackUp, node distance=2cm] (Neuron1) { 
		%
		$\begin{array}{c}
			\text{Neuron } 1\\  (\xi_1, \Delta, \alpha)
		\end{array}$
	};
	\node [blockW, right of=stateFeedbackDown, node distance=2cm] (Neuron2) { 
		$\begin{array}{c}
			\text{Neuron } 2\\  (\xi_2, \Delta, \alpha)
		\end{array}$
	};
	
	\draw [draw,-] (stateFeedback) -- node [pos=0.5]{$y$} (stateFeedbackRight);
	\draw [draw,-] (stateFeedbackRight) --  (stateFeedbackUp);
	\draw [draw,-] (stateFeedbackRight) --  (stateFeedbackDown);
	\draw [draw,->] (stateFeedbackUp) -- node {} (Neuron1);
	\draw [draw,->] (stateFeedbackDown) -- node {} (Neuron2);
	
%
	
	\node [input, right of= Neuron1, node distance=4cm] (Neuron1End){};
	\node [input, right of= Neuron2, node distance=4cm] (Neuron2End){};
	\node [blockW, right of=Neuron1, node distance=3.2cm] (Neuron1Gain) { 
		$-1$
	};
	
	\draw [draw,-] (Neuron2) -- node {} (Neuron2End);
	\draw [draw,->] (Neuron1) -- node {} (Neuron1Gain);
	\draw [draw,-] (Neuron1Gain) -- node {} (Neuron1End);
	
	\node [input, below of= Neuron1End, node distance=0.9cm] (Neuron1Point){};
	\node [input, above of= Neuron2End, node distance=0.9cm] (Neuron2Point){};
	\node [input, below of= Neuron1Point, node distance=0.1cm] (NeuronCircle){};
	\draw [draw,->] (Neuron1End) -- node {} (Neuron1Point);
	\draw [draw,->] (Neuron2End) -- node {} (Neuron2Point);
	
	\draw [fill=white] (NeuronCircle) circle (0.1cm);
	
	\node [input, right of= NeuronCircle, node distance=0.1cm] (NeuronCircleRight){};
	
	\node [blockW, right of=NeuronCircle, node distance=2.5cm] (system) { 
		$\begin{array}{c}
			\text{Plant }
		\end{array}$
	};
	
	\draw [draw,->] (NeuronCircleRight) -- node [pos=0.8]{$u$} (system);
	
	\node [input, right of= system, node distance=2cm] (SystemOutput){};
	\node [input, right of= system, node distance=1.5cm] (SystemFeedback){};
	\node [input, below of= SystemFeedback, node distance=1.8cm] (SystemFeedbackBelow){};
	\node [input, below of= stateFeedback, node distance=1.8cm] (stateFeedbackBelow){};
	
	\draw [draw,->] (system) -- node [pos=0.5]{$y$} (SystemOutput);
	\draw [draw,-] (SystemFeedback) --  (SystemFeedbackBelow);
	\draw [draw,-] (stateFeedback) --  (stateFeedbackBelow);
	\draw [draw,-] (SystemFeedbackBelow) --  (stateFeedbackBelow);

	\node [input, right of= Neuron1, node distance=1.5cm] (Neuron1SpikeBeginDown){};
	\node [input, above of= Neuron1SpikeBeginDown, node distance=0.2cm] (Neuron1SpikeBegin){};
	\node [input, above of= Neuron1SpikeBegin, node distance=0.6cm] (Neuron1Spike1Up){};
	\node [input, right of= Neuron1SpikeBegin, node distance=0.1cm] (Neuron1Spike2Down){};
	\node [input, left of= Neuron1SpikeBegin, node distance=0.2cm] (Neuron1SpikeBeginLeft){};
	\node [input, above of= Neuron1Spike2Down, node distance=0.6cm] (Neuron1Spike2Up){};
	\node [input, right of= Neuron1Spike2Down, node distance=0.4cm] (Neuron1Spike3Down){};
	\node [input, above of= Neuron1Spike3Down, node distance=0.6cm] (Neuron1Spike3Up){};
	\node [input, right of= Neuron1Spike3Down, node distance=0.3cm] (Neuron1Spike4Down){};
	\node [input, above of= Neuron1Spike4Down, node distance=0.6cm] (Neuron1Spike4Up){};
	\node [input, right of= Neuron1Spike3Down, node distance=0.2cm] (Neuron1Spike4Down){};
	\node [input, above of= Neuron1Spike4Down, node distance=0.6cm] (Neuron1Spike4Up){};
	\node [input, right of= Neuron1Spike4Down, node distance=0.2cm] (Neuron1SpikeFinal){};
	
	\draw [draw,-] (Neuron1SpikeBegin) --  (Neuron1SpikeBeginLeft);
	\draw [draw,-] (Neuron1SpikeBegin) --  (Neuron1Spike1Up);
	\draw [draw,-] (Neuron1SpikeBegin) --  (Neuron1Spike2Down);
	\draw [draw,-] (Neuron1Spike2Down) --  (Neuron1Spike2Up);
	\draw [draw,-] (Neuron1Spike2Down) --  (Neuron1Spike3Down);
	\draw [draw,-] (Neuron1Spike3Down) --  (Neuron1Spike3Up);
	\draw [draw,-] (Neuron1Spike3Down) --  (Neuron1Spike4Down);
	\draw [draw,-] (Neuron1Spike4Down) --  (Neuron1Spike4Up);
	\draw [draw,-] (Neuron1Spike4Down) --  (Neuron1SpikeFinal);
	
	\node [input, right of= Neuron2, node distance=1.5cm] (Neuron2SpikeBeginDown){};
	\node [input, above of= Neuron2SpikeBeginDown, node distance=0.2cm] (Neuron2SpikeBegin){};
	\node [input, above of= Neuron2SpikeBegin, node distance=0.6cm] (Neuron2Spike1Up){};
	\node [input, right of= Neuron2SpikeBegin, node distance=0.3cm] (Neuron2Spike2Down){};
	\node [input, left of= Neuron2SpikeBegin, node distance=0.1cm] (Neuron2SpikeBeginLeft){};
	\node [input, above of= Neuron2Spike2Down, node distance=0.6cm] (Neuron2Spike2Up){};
	\node [input, right of= Neuron2Spike2Down, node distance=0.2cm] (Neuron2Spike3Down){};
	\node [input, above of= Neuron2Spike3Down, node distance=0.6cm] (Neuron2Spike3Up){};
	\node [input, right of= Neuron2Spike3Down, node distance=0.2cm] (Neuron2Spike4Down){};
	\node [input, above of= Neuron2Spike4Down, node distance=0.6cm] (Neuron2Spike4Up){};
	\node [input, right of= Neuron2Spike3Down, node distance=0.3cm] (Neuron2Spike4Down){};
	\node [input, above of= Neuron2Spike4Down, node distance=0.6cm] (Neuron2Spike4Up){};
	\node [input, right of= Neuron2Spike4Down, node distance=0.2cm] (Neuron2SpikeFinal){};
	
	\draw [draw,-] (Neuron2SpikeBegin) --  (Neuron2SpikeBeginLeft);
	\draw [draw,-] (Neuron2SpikeBegin) --  (Neuron2Spike1Up);
	\draw [draw,-] (Neuron2SpikeBegin) --  (Neuron2Spike2Down);
	\draw [draw,-] (Neuron2Spike2Down) --  (Neuron2Spike2Up);
	\draw [draw,-] (Neuron2Spike2Down) --  (Neuron2Spike3Down);
	\draw [draw,-] (Neuron2Spike3Down) --  (Neuron2Spike3Up);
	\draw [draw,-] (Neuron2Spike3Down) --  (Neuron2Spike4Down);
	\draw [draw,-] (Neuron2Spike4Down) --  (Neuron2Spike4Up);
	\draw [draw,-] (Neuron2Spike4Down) --  (Neuron2SpikeFinal);

	\node [input, right of= NeuronCircle, node distance=0.35cm] (Neuron1SpikeBeginDown_tot){};
	\node [input, above of= Neuron1SpikeBeginDown_tot, node distance=0.8cm] (Neuron1SpikeBegin_tot){};
	\node [input, above of= Neuron1SpikeBegin_tot, node distance=-0.6cm] (Neuron1Spike1Up_tot){};
	\node [input, right of= Neuron1SpikeBegin_tot, node distance=0.1cm] (Neuron1Spike2Down_tot){};
	\node [input, left of= Neuron1SpikeBegin_tot, node distance=0.2cm] (Neuron1SpikeBeginLeft_tot){};
	\node [input, above of= Neuron1Spike2Down_tot, node distance=-0.6cm] (Neuron1Spike2Up_tot){};
	\node [input, right of= Neuron1Spike2Down_tot, node distance=0.4cm] (Neuron1Spike3Down_tot){};
	\node [input, above of= Neuron1Spike3Down_tot, node distance=-0.6cm] (Neuron1Spike3Up_tot){};
	\node [input, right of= Neuron1Spike3Down_tot, node distance=0.3cm] (Neuron1Spike4Down_tot){};
	\node [input, above of= Neuron1Spike4Down_tot, node distance=-0.6cm] (Neuron1Spike4Up_tot){};
	\node [input, right of= Neuron1Spike3Down_tot, node distance=0.2cm] (Neuron1Spike4Down_tot){};
	\node [input, above of= Neuron1Spike4Down_tot, node distance=-0.6cm] (Neuron1Spike4Up_tot){};
	\node [input, right of= Neuron1Spike4Down_tot, node distance=0.2cm] (Neuron1SpikeFinal_tot){};
	
	\draw [draw,-] (Neuron1SpikeBegin_tot) --  (Neuron1SpikeBeginLeft_tot);
	\draw [draw,-] (Neuron1SpikeBegin_tot) --  (Neuron1Spike1Up_tot);
	\draw [draw,-] (Neuron1SpikeBegin_tot) --  (Neuron1Spike2Down_tot);
	\draw [draw,-] (Neuron1Spike2Down_tot) --  (Neuron1Spike2Up_tot);
	\draw [draw,-] (Neuron1Spike2Down_tot) --  (Neuron1Spike3Down_tot);
	\draw [draw,-] (Neuron1Spike3Down_tot) --  (Neuron1Spike3Up_tot);
	\draw [draw,-] (Neuron1Spike3Down_tot) --  (Neuron1Spike4Down_tot);
	\draw [draw,-] (Neuron1Spike4Down_tot) --  (Neuron1Spike4Up_tot);
	\draw [draw,-] (Neuron1Spike4Down_tot) --  (Neuron1SpikeFinal_tot);
	
	\node [input, right of= NeuronCircle, node distance=0.3cm] (Neuron2SpikeBeginDown_tot){};
	\node [input, above of= Neuron2SpikeBeginDown_tot, node distance=0.8cm] (Neuron2SpikeBegin_tot){};
	\node [input, below of= Neuron2SpikeBegin_tot, node distance=-0.6cm] (Neuron2Spike1Up_tot){};
	\node [input, right of= Neuron2SpikeBegin_tot, node distance=0.3cm] (Neuron2Spike2Down_tot){};
	\node [input, left of= Neuron2SpikeBegin_tot, node distance=0.1cm] (Neuron2SpikeBeginLeft_tot){};
	\node [input, below of= Neuron2Spike2Down_tot, node distance=-0.6cm] (Neuron2Spike2Up_tot){};
	\node [input, right of= Neuron2Spike2Down_tot, node distance=0.2cm] (Neuron2Spike3Down_tot){};
	\node [input, below of= Neuron2Spike3Down_tot, node distance=-0.6cm] (Neuron2Spike3Up_tot){};
	\node [input, right of= Neuron2Spike3Down_tot, node distance=0.2cm] (Neuron2Spike4Down_tot){};
	\node [input, below of= Neuron2Spike4Down_tot, node distance=-0.6cm] (Neuron2Spike4Up_tot){};
	\node [input, right of= Neuron2Spike3Down_tot, node distance=0.3cm] (Neuron2Spike4Down_tot){};
	\node [input, below of= Neuron2Spike4Down_tot, node distance=-0.6cm] (Neuron2Spike4Up_tot){};
	\node [input, right of= Neuron2Spike4Down_tot, node distance=0.2cm] (Neuron2SpikeFinal_tot){};
	
	\draw [draw,-] (Neuron2SpikeBegin_tot) --  (Neuron2Spike1Up_tot);
	\draw [draw,-] (Neuron2Spike2Down_tot) --  (Neuron2Spike2Up_tot);
	\draw [draw,-] (Neuron2Spike3Down_tot) --  (Neuron2Spike3Up_tot);
	\draw [draw,-] (Neuron2Spike4Down_tot) --  (Neuron2Spike4Up_tot);
\end{tikzpicture}
\end{center}
\caption{Block diagram representing the system architecture }
\label{Fig:blockDiagram_op2}
\end{figure}
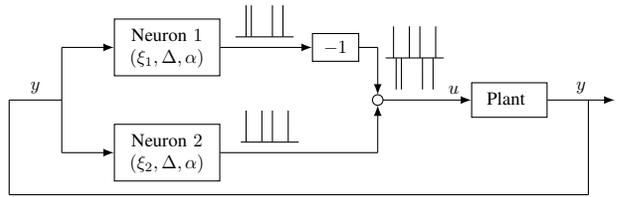

Consider the linear time-invariant system 
\begin{equation}
\begin{aligned}
&\dot x = Ax + Bu \\
&
 y = Cx, 
\end{aligned}
\label{eq:system}
\end{equation}
where 
$x \in \R^{n_x}$  is the state, $y \in \R$ is the output, 
$u \in \R$ the spiking control input and $n_x\in \Z_{> 0}$. 
  In particular, we consider the setting shown in Fig.~\ref{Fig:blockDiagram_op2}, where the input $u$ is generated by a neuromorphic controller consisting of two variables, $\xi_\ell \in \R_{\geq0}$, $\ell \in \{1,2\}$, whose dynamics is inspired by the leaky integrate-and-fire neuron model \cite{lapicque1907recherches,abbott1999lapicque, izhikevich2010hybrid}, \cite[Chapter 4]{gerstner2002spiking}. In particular, $\xi_\ell \in \R_{\geq0}$, $\ell \in \{1,2\}$, represents the membrane potential of the neuron, whose continuous-time dynamics is a first-order differential equation affected by a non-negative current input, and it is then reset whenever it reaches a positive threshold.  Note that, as explained in \cite{gerstner2002spiking}, the integrate-and-fire neuron model generates spiking times only when fed with non-negative inputs and it is insensitive to negative inputs. 
	%
The emitted spike, considered in the form of a Dirac delta pulse $\alpha \delta(t-t_j)$, with  $\alpha \in \R_{> 0}$ a given and constant spike amplitude, $t \in \R_{\geq 0}$ and $t_j$ $j \in \Z_{\geq 0}$ the considered spiking time, is absorbed by the actuator and leads to the reset of the plant state $x \in \R^{n_x}$,  which we will make more precise below, see \eqref{eq:thresholdJump}. In our neuromorphic controller design, the input to the neurons is the continuous-time measured output $y$ of system \eqref{eq:system}, which, depending on its sign, influences either $\xi_1$ or $\xi_2$. Indeed, since the measured output $y \in \R$ and the neuron dynamics is affected only by non-negative inputs, in the controller design we need one neuron sensitive to $y$ and one to $-y$. 

Inspired by the leaky integrate-and-fire neuron model \cite{lapicque1907recherches,abbott1999lapicque, izhikevich2010hybrid}, \cite[Chapter 4]{gerstner2002spiking}, we thus consider the following dynamics for the neuron-like variables  $\xi_\ell$, $\ell \in \{1,2\}$, between two control actions, 
%
\begin{equation}
\begin{aligned}
&\dot \xi_1 = -\mu \xi_1 +  \max(0,y), \quad 
\dot \xi_2 = -\mu \xi_2 + \max(0,-y), 
\end{aligned}
\label{eq:thresholdFlow}
\end{equation}
with $\mu \in \R_{\geq 0}$ a ``leaking" parameter of the neurons.
%
%
Moreover, the neuron-like variables $\xi_\ell$, $\ell \in \{1,2\}$, trigger spikes
when there exists $\ell \in \{1,2\}$ such that the condition
\begin{equation}
\xi_\ell \geq \Delta
\label{eq:triggeringRule}
\end{equation}
is satisfied, where $\Delta \in \R_{> 0}$ is, again, a neuron parameter. 
For ease of exposition, we take $\mu$ and $\Delta$ the same for both neurons, but they can also be taken differently, see Remark~\ref{Rem:AsymmetricNeurons} below.
%
Inspired by the leaky integrate-and-fire neuron model, where the membrane potential is reset whenever it reaches a threshold, we reset the variable $\xi_\ell$, $\ell \in \{1,2\}$, when it triggers a spiking control action, while it is not modified when the other neuron-like variable reaches the threshold,  i.e., when neuron $\ell \in \{1,2\}$ satisfies \eqref{eq:triggeringRule}, we have, in terms of the hybrid system notation in Section \ref{Notation},
\begin{equation}
\xi_\ell^+ = 0
\label{eq:thresholdJump}
\ \text{ and } \
\xi_{3-\ell}^+ = \xi_{3-\ell}. 
\end{equation}
%
It is important to notice that the leaky integrate-and-fire neuron model does not include a spike generation mechanism, but, as explained in \cite{izhikevich2010hybrid}, it is only a threshold mechanism and therefore does not intrinsically produce a spike. 
To represent the neuron behavior more accurately, in our design we add a spike generation mechanism to the leaky integrate-and-fire neuron model. Indeed, whenever $\xi_\ell$, $\ell \in \{1,2\}$, reaches the threshold $\Delta$, a spiking control action, which can be interpreted by the release of charge of the membrane potential affecting the connected system, is fired, whose sign
depends on which neuron-like variable reaches $\Delta$.
In particular, when $\xi_1$ reaches the threshold $\Delta$, the spiking control action at the spiking time $t_j$, $j \in \Z_{\geq 0}$, is given by, with some abuse of notation, $-\alpha \delta (t-t_j)$, with $t \in \R_{\geq 0}$, which results, in terms of the hybrid system notation in Section \ref{Notation}, 
\begin{equation}
x^+ = x - B\alpha. 
\label{eq:systemJump1}
\end{equation}
Similarly, when $\xi_2$ reaches the threshold $\Delta$, the spiking control action at the spiking time $t_j$, $j \in \Z_{\geq 0}$, is given by, again with some abuse of notation, $\alpha \delta (t-t_j)$, with $t \in \R_{\geq 0}$, and thus 
\begin{equation}
x^+ = x + B\alpha.
\label{eq:systemJump2}
\end{equation}
Note that, equations \eqref{eq:systemJump1}-\eqref{eq:systemJump2} assume that the actuator can cope with spiking control signals, which is reasonable in some applications, see e.g., \cite{van2018organic,krauhausen2021organic}. Interestingly, the number and type of neuromorphic devices, such as actuators and sensors, which are important for the real-life implementation of true neuromorphic controllers, such as \eqref{eq:thresholdFlow}, \eqref{eq:triggeringRule}, leading to \eqref{eq:systemJump1}, \eqref{eq:systemJump2}, are growing in the recent years due the engineering community's interests in developing neuromorphic closed-loop systems. A prominent example is the event-based camera \cite{singh2016stabilization, singh2018regulation}.

Between two spiking control instants, i.e., when for all $\ell \in \{1,2\}$, $\xi_\ell \leq \Delta$, $u = 0$ and thus the system state evolves in open loop according to 
\begin{equation}
\dot x = Ax.
\label{eq:systemFlow}
\end{equation}

To capture all of the above, we define the overall state as $q:= (x, \xi_1, \xi_2) \in \R^{n_x +2}$ and we obtain the hybrid system 
\begin{equation}
\left\lbrace \
\begin{aligned}
\dot{q} &= F(q), \ \ \ \ \ &&q \in \mathcal{C}\\
q^{+} &\in G(q), \ \ \ \ \ &&q \in \mathcal{D},
\end{aligned}
\right.
\label{eq:HybridSystem}
\end{equation} 
where the flow map $F$ is defined, for any $q \in \mathcal{C}$, from \eqref{eq:thresholdFlow} and \eqref{eq:systemFlow}, 
\begin{equation}
F(q):= (Ax, -\mu \xi_1 +  \max(0,y), -\mu \xi_2 +  \max(0, -y)). 
\label{eq:flowHybridSystem}
\end{equation}
The flow set $\mathcal{C}$ in \eqref{eq:HybridSystem} is defined as
\begin{equation}
\mathcal{C} :=\mathcal{C}_1\cap \mathcal{C}_2
\label{eq:flowSet_bis}
\end{equation}
with
\begin{equation}
\mathcal{C}_\ell := \left\{q \in \R^{n_x +2}: \xi_\ell \leq \Delta\right\}, \quad \ell \in \{1,2\}.
\label{eq:flowSet_i}
\end{equation}

The jump set $\mathcal{D}$ in \eqref{eq:HybridSystem} is defined as, from \eqref{eq:triggeringRule}, 
\begin{equation}
\mathcal{D} :=\mathcal{D}_1\cup \mathcal{D}_2
\label{eq:jumpSet}
\end{equation}
with 
\begin{equation}
\mathcal{D}_\ell :=\left\{q \in \R^{n_x+2}: \xi_\ell \geq \Delta\right\}, \quad \ell \in \{1,2\}.
\label{eq:jumpSet_i}
\end{equation}

The jump map $G$ in \eqref{eq:HybridSystem} is defined as, for any $q \in \mathcal{D}$, from \eqref{eq:systemJump1}-\eqref{eq:thresholdJump}, 
\begin{equation}
G(q):= G_1(q) \cup G_2(q), 
\label{eq:jumpMap}
\end{equation}
with 
\begin{equation}
G_1(q):= 	\left\lbrace 
\begin{aligned}
&\left\{\left(
\begin{array}{c}
x- B\alpha\\
0\\
\xi_2\\
\end{array}
\right) \right\} 
\ \ \  &&q \in \mathcal{D}_1\\
& \qquad \quad \! \! \emptyset &&q \notin \mathcal{D}_1,
\end{aligned}
\right. 
\label{eq:JumpMap_1}
\end{equation}
and
\begin{equation}
G_2(q):= 	\left\lbrace 
\begin{aligned}
&\left\{\left(
\begin{array}{c}
x+ B\alpha\\
\xi_1\\
0\\
\end{array}
\right) \right\}  
\ \ \  &&q \in \mathcal{D}_2\\
& \qquad \quad \! \! \emptyset &&q \notin \mathcal{D}_2.
\end{aligned}
\right. 
\label{eq:JumpMap_2}
\end{equation}

\begin{rem}
	The current formulation includes a pair of neuron-like variables. We can envision a model with multiple pairs of neuron-like variables, with possibly a control input $u$ with dimension greater than $1$. This is the subject of future research. 
\end{rem}

	Now that we have proposed the neuromorphic controller and provided a suitable modeling framework, it is our
	 objective to derive conditions on the neuron parameters $\alpha, \mu, \Delta$ to guarantee a stability property of system~\eqref{eq:HybridSystem}. In the next section we present results for the case when \eqref{eq:system} is a scalar state linear time-invariant system.

\section{Main analysis and design}\label{MainResult}
The goal of this section is to prove that the proposed neuromorphic controller guarantees a stability property for system \eqref{eq:HybridSystem} in the particular case of single-state linear time-invariant systems. In particular, in the next theorem we show that system \eqref{eq:HybridSystem} with $A =a \in\R_{> 0}$, $B = C = 1$, satisfies a  semi-global practical stability property, and we ensure the existence of a strictly positive dwell-time, thereby ruling out the Zeno phenomenon. 
We will also show how the control parameters (leaking rate $\mu$, firing threshold $\Delta$ and spike amplitude $\alpha$) and plant parameter $a$ relate to the region of attraction, the ultimate bound and the dwell-time guarantees.
Note that, the case $B=C=1$ is considered in Theorem~\ref{Thm:StabilityTheorem} without loss of generality and the results hold \emph{mutatis mutandi} for any $B,C \in \R_{>0}$. The case $B=C=1$ is considered just to simplify notations.


\begin{thm}
Consider system \eqref{eq:HybridSystem} with $n_x = 1$ and $A = a \in \R_{>0}$, $B = C = 1$. Select $\rho \in (0,1)$. Define $\Psi:= \displaystyle \frac{\rho  + 1}{(\rho + 1)^2 -1}\alpha$. 
Then for all $\displaystyle \Delta \in \left(0, \frac{\rho\alpha}{\mu +a}\right]$, and all $\sigma \in \left[\displaystyle \frac{(\rho + 1)^2 -1}{(\rho + 1)^2},1\right)$, 
any solution $q$ of the controlled system~\eqref{eq:HybridSystem} such that $\displaystyle |x(0,0)| \leq \sigma\Psi$ and $\xi_1(0,0) = \xi_2(0,0) = 0$ 
satisfies the following properties: 
\begin{enumerate}[label=(\roman*)] 
\item 
For all $(t,j) \in \dom q$ 
\begin{equation}
|x(t,j)| \leq  \gamma^j |x(0, 0)| + 2\alpha,
\label{eq:Th1_3}
\end{equation}
where $\displaystyle \gamma :=  (1 - (1-\sigma)\left(\rho +1\right)^2)^{\frac{1}{2}} \in (0,1)$. 
\item The solution $q$ has a minimum dwell-time $\displaystyle \tau:= \frac{\Delta}{\Upsilon}\in \R_{>0}$, with $\Upsilon:= \Psi + 2\alpha \in \R_{> 0}$,  
i.e., for all $(s,i), (t,j) \in \dom q$ with $s+i \leq t+j$ we have $\displaystyle j-i \leq \frac{t-s}{\tau} + 1$. 
\end{enumerate}
\label{Thm:StabilityTheorem}
\end{thm}

Theorem~\ref{Thm:StabilityTheorem}, whose proof is given in the Appendix, 
provides conditions on the control parameters $\alpha$, $\Delta$, $\mu$, to ensure that the proposed neuromorphic controller guarantees a semi-global practical stability property for system \eqref{eq:HybridSystem}. Indeed, item~\textit{(i)} guarantees that, when the system initial state is such that $|x(0,0)| \leq \sigma \Psi$, with $\Psi$ that can be arbitrarily large (by taking $\rho$ close to zero), it converges to a neighborhood of the origin, whose size is given by $2\alpha$, as the time grows. 
Since the controller operates via fixed amplitude spikes, it is not possible to obtain an asymptotic stability property for the closed-loop system, and, thus, proving a practical one as in Theorem~\ref{Thm:StabilityTheorem}, is the best we can achieve.
Note that, since the solution to system \eqref{eq:system} is diverging between spikes, because of $a>0$, and the magnitude $\alpha$ of the proposed controller is given and fixed, the stability result provided in this theorem is not trivial. Indeed, the time occurrence of the spiking control actions needs to be sufficiently fast to guarantee that each continuous-time divergent behavior between two consecutive control actions is compensated by the discrete spike. This is reflected by the $a$, $\alpha$ and $\mu$ dependence of the upperbound on the firing threshold $\Delta$. Consequently, for a given spike amplitude, the further the state $x$ is from the origin, the faster the spikes must be generated, which provides a trade-off between region of attraction and the dwell-time, as reflected by the conditions on the parameters in Theorem~\ref{Thm:StabilityTheorem}.
Note that, it is possible to select the design parameters so that the region of attraction can be arbitrarily large.  Indeed, as can be seen from the conditions in Theorem~\ref{Thm:StabilityTheorem}, the smaller $\rho \in (0,1)$ is chosen, the larger $\Psi$ will be. However, the smaller $\rho$ is selected, the smaller is the upperbound on the firing threshold $\Delta$ that guarantees the stability property. Thus, in turn, as stated in item \textit{(ii)}, decreases the guaranteed dwell-time. Similarly, also the parameter $\mu \in \R_{\geq 0}$ has an impact on the threshold $\Delta$. Indeed, the larger the $\mu$ (the ``leakier" the neuronal dynamics are), the smaller the upperbound on $\Delta$ is, which decreases the guaranteed dwell-time.
However, for any choice of the design parameters satisfying the conditions in Theorem \ref{Thm:StabilityTheorem}, item~\textit{(ii)} guarantees the existence of a strictly positive dwell-time, which implies that the Zeno phenomenon cannot occur. 
In addition, item~\textit{(i)} provides an upperbound on the guaranteed convergence speed. In particular, the smaller the $\sigma$, that is, the farther the initial state is from the border of the region of attraction, the larger $\gamma$ is, which affects the convergence speed.  
Finally, the result holds for any spike amplitude $\alpha \in \R_{> 0}$. From Theorem~\ref{Thm:StabilityTheorem} we have that the smaller the $\alpha$, the smaller the ultimate bound towards which the state converges is. However, the smaller  the $\alpha$, the smaller the region of attraction and the upperbound on the threshold $\Delta$ are. 


Conversely, from a controller design perspective, Theorem~\ref{Thm:StabilityTheorem} guarantees that it is possible to obtain any desirable region of attraction $\Psi \in\R_{> 0}$ and ultimate bound $2\alpha\in \R_{> 0}$ for the closed-loop system for any given $\mu$, if there is freedom in the selection of the parameters $\alpha$ and $\Delta$ of neuron-like variables $\xi_\ell$, $\ell \in \{1,2\}$. Indeed, the ultimate bound is directly related to the amplitude of the spikes $\alpha$. Moreover, given $\alpha$ and the desired $\Psi$, it is always possible to find $\rho \in (0,1)$ such that $\Psi = \frac{\rho +1}{(\rho+1)^2 -1}\alpha$ is satisfied. Then, for any given $\mu$, by selecting $\Delta$ such that $\Delta \in \left(0, \frac{\rho\alpha}{\mu +a}\right]$ holds, the practical stability property with the desired ultimate bound $2\alpha$ and region of attraction $\Psi$ is guaranteed. 
%


From \eqref{eq:thresholdFlow}-\eqref{eq:triggeringRule}, \eqref{eq:thresholdJump}, and since $\xi_\ell (0,0) = 0$, $\ell \in \{1,2\}$, from the conditions in Theorem \ref{Thm:StabilityTheorem}, we have that the neuron-like variables are always bounded, namely $\xi_\ell (t,j) \leq \Delta$, $\ell \in \{1,2\}$, for all $(t,j) \in \dom q$. 
\begin{rem} 
	In Theorem~\ref{Thm:StabilityTheorem} the neuron-like variables $\xi_1$ and $\xi_2$ are initialized at $\xi_1(0,0) = \xi_2(0,0) = 0$. This is just a design choice that simplifies the proof of Theorem~\ref{Thm:StabilityTheorem}, but any other choice for the initial condition would not compromise the stability result in item \textit{(i)}. However, the dwell-time property in item \textit{(ii)} would hold only after the first jump of each neuron-like variable. 
\end{rem}

\begin{rem}
	Theorem~\ref{Thm:StabilityTheorem} provides a stability results for system~\eqref{eq:HybridSystem} with $A= a > 0$, $B=C=1$. 
	Similar results hold \emph{mutatis mutandis} for system \eqref{eq:HybridSystem} with $A= a \leq 0$, $B,C \in \R_{> 0}$. Moreover, when $a <0$ an asymptotic stability property can be proved under some conditions. 
\end{rem}

\begin{rem}
	The neuron parameters $\alpha, \mu, \Delta$ are assumed to be the same in the two neuron-like variables $\xi_\ell$, $\ell \in \{1,2\}$ in~\eqref{eq:HybridSystem}. The proposed model and results hold \emph{mutatis mutandis} when considering different neuron parameters for the two variables. In particular, we can consider asymmetric spike amplitude, namely $u \in \{0, +\alpha_1, -\alpha_2\}$, with $\alpha_1, \alpha_2 \in \R_{> 0}$, different leaking parameters $\mu_1, \mu_2 \in \R_{\geq 0}$ and thresholds $\Delta_1, \Delta_2 \in \R_{> 0}$.  More details will be provided in future work. 
\label{Rem:AsymmetricNeurons}
\end{rem}


\section{Numerical example}\label{Example}
\begin{figure}
	\centering
	\includegraphics[scale=0.385]{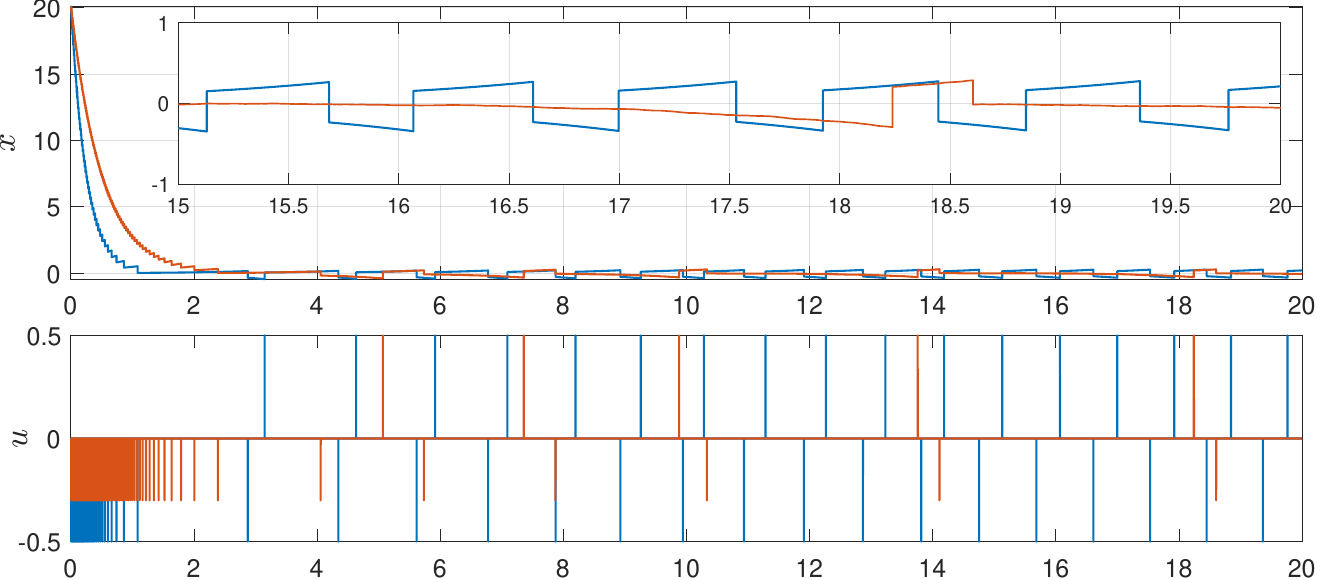}
	\caption{State $x$ and input $u$ when $\alpha = 0.5$, $\mu = 0.5$, $\Delta = 0.1$ (blue) and  $\alpha_1 = 0.3$, $\alpha_2 = 0.5$, $\mu_1 = 0.2$, $\mu_2 = 0.5$, $\Delta_1 = 0.1$, $\Delta_2 = 0.2$ with noise and disturbances (red).}
	\label{fig:ExampleSymNoNoiseAndAsymWithNoise}
\end{figure}

We consider system \eqref{eq:system} with $A = 1$, $B = 1$, $C=1$ and we apply the proposed neuromorphic spiking controller, where the neuron parameters are $\alpha = 0.5$, $\mu = 0.5$, $\Delta = 0.1$, which satisfy the conditions of Theorem~\ref{Thm:StabilityTheorem} for any $\rho \in (0,1)$. The simulation results for the system state $x$ and the corresponding spiking control input $u$, obtained with the initial conditions $x(0,0) = 20$, $\xi_1(0,0) = \xi_2(0,0) = 0$, are shown in Fig.~\ref{fig:ExampleSymNoNoiseAndAsymWithNoise} in blue. A zoom of the ultimate bound of the state convergence is also shown in Fig.~\ref{fig:ExampleSymNoNoiseAndAsymWithNoise}.

Fig.~\ref{fig:ExampleSymNoNoiseAndAsymWithNoise} shows that the proposed neuromorphic controller satisfies a practical stability property of the closed-loop system. The ultimate bound obtained in simulations is $0.375$, which is smaller than the guaranteed one from Theorem \ref{Thm:StabilityTheorem}, equal to $2\alpha = 1$. In addition, the minimum time between two spikes obtained in the whole simulation is $0.005$ and, it is equal $0.341$ when considering only the steady-state. 

To investigate the effect of disturbances and measurement noises,
we also consider the same system parameters with modified  
state dynamics $\dot x = x + v$, where $v \in \R$ is an external disturbance, and modified output equation $y = x +w$, where $w \in \R$ is the measurement noise. Both $v$ and $w$ used in the simulation are piecewise linear signals generated from a random sequence of points, occurring at intervals of $0.01$~s, with amplitude ranging between $-0.1$ and $0.1$ (uniform distribution) and with a linear interpolation applied between any two consecutive points. Moreover, we also consider different neuron parameters for the two neuron-like variables in \eqref{eq:thresholdFlow}-\eqref{eq:thresholdJump}. In particular, we select $\alpha_1 = 0.3$, $\alpha_2 = 0.5$, $\mu_1 = 0.2$, $\mu_2 = 0.5$, $\Delta_1 = 0.1$ and $\Delta_2 = 0.2$. The simulation results obtained in this setting are shown Fig.~\ref{fig:ExampleSymNoNoiseAndAsymWithNoise} in red, 
%
which illustrates that the proposed neuromorphic controller is robust to measurement noise, disturbances and differences on the neuron parameters $\alpha_\ell$, $\mu_\ell$ and $\Delta_\ell$, $\ell \in \{1,2\}$. 
A formal analysis of the asymmetric case and of the effect of disturbances and measurement noises is left for future work. 

\section{Conclusions}\label{Conclusions}
We have presented a hybrid system framework to formally analyze a neuromorphic controller in closed-loop with single-input single-output linear time-invariant systems. 
The controller is based on two neuron-like variables, whose dynamics is inspired by the leaky integrate-and-fire neuron model. 
Using this hybrid system model, 
we have proved that the neuromorphic controller can guarantee a practical stability property for the system state in the particular case of single-state linear time-invariant systems. Moreover, we have related this property to the neuron parameters. Finally, a numerical example confirms the obtained analysis and design.  

We believe that the presented framework is the starting point for many interesting future work directions among which the generalization of the results to higher-order linear time-invariant systems and/or nonlinear systems. This might require multiple pairs of neurons in the controller, and therefore new designs and structures. We would also like to include measurement noise and disturbances in the system model and provide a formal analysis for the resulting closed-loop system. 
Moreover, in future work we plan to relax the assumption that the parameters of the two neuron-like variables are the same, in the sense that we would like to consider possible asymmetric spike amplitudes, leaking parameters and thresholds for the different neurons. Considering the case where the controller receives spiking measurements from neuromorphic sensors, and not continuous-time ones, is an additional challenging research direction. 
Finally, it would be interesting to run experiments on physical systems implementing the proposed neuromorphic controller. 

\appendix
\noindent\textbf{Proof of Theorem~\ref{Thm:StabilityTheorem}.} 
Let all conditions of Theorem~\ref{Thm:StabilityTheorem} hold and consider the Lyapunov function candidate $V(x) := x^2$, for $x \in \R$. 

Let $q \in \mathcal{C}$, in view of \eqref{eq:flowHybridSystem}, 
\begin{equation}
	\begin{aligned}
		\left\langle \nabla V(x), ax \right\rangle &= 2ax^2 
		= 2aV(x). 
	\end{aligned}
	\label{eq:EqProofTh1_flow}
\end{equation}

Let $q \in \mathcal{D}_1$, from \eqref{eq:JumpMap_1} we have
\begin{equation}
	V(x^+) = (x- \alpha)^2. 
	\label{eq:EqProofTh1_1}
\end{equation}
%
Similarly, let $q \in \mathcal{D}_2$, from \eqref{eq:JumpMap_2} we have
\begin{equation}
	V(x^+) = (x+ \alpha)^2. 
	\label{eq:EqProofTh1_2}
\end{equation}

Let $q$ be a solution to system \eqref{eq:HybridSystem} with $|x(0,0)| \leq \sigma\Psi$. Pick any $(t,j) \in \dom q$ and let $0 = t_0 \leq t_1 \leq \dots \leq t_{j+1} = t$ satisfy $\dom q \cap ([0,t]\times \{0,1,\dots,j\}) = \bigcup_{i=0}^j [t_i, t_{i+1}] \times \{i\}$. 
Note that, $t_{j+1} = t$ is not necessarily a jump time but $t_1, \dots, t_j$ are jump times.
For each $i \in \{0,1,\dots,j\}$ and all $s \in [t_i, t_{i+1}]$, $q(s,i) \in \mathcal{C}$ and, from \eqref{eq:EqProofTh1_flow} we have
	$V(x(s,i)) = e^{2a(s-t_i)}V(x(t_i,i))$, 
which implies, for each $i \in \{0,1,\dots,j\}$, 
\begin{equation}
	V(x(t_{i+1},i)) = e^{2a(t_{i+1}-t_i)}V(x(t_i,i)). 
	\label{eq:EqProofTh1_3} 
\end{equation}
Similarly, for each $i \in \{0,1,\dots,j\}$ and all $s \in [t_i, t_{i+1}]$, \eqref{eq:systemFlow} implies
\begin{equation}
	x(s,i) = e^{a(s-t_i)}x(t_i,i). 
	\label{eq:StateFlow_proofTh1}
\end{equation}
Thus, for each $i \in \{0,1,\dots,j\}$ and  all $s \in [t_i, t_{i+1}]$, $\sign{x(s,i)} = \sign{x(t_i,i)}$. 
%
%
Moreover for each $i \in \{0,1,\dots,j\}$ and all $s \in [t_i, t_{i+1}]$, from \eqref{eq:thresholdFlow} we have
\begin{equation}
	\begin{aligned}
	\xi_1(s,i) = & \ e^{-\mu (s-t_i)}\xi_1(t_i,i) \\
	&+ \int_{t_i}^{s} e^{-\mu (s-\tilde{s})} \max (0,x(\tilde{s},i)) d\tilde{s}
	\end{aligned}
	\label{eq:thresholdTime}
\end{equation} 
and
\begin{equation}
	\begin{aligned}
	\xi_2(s,i) =  & \ e^{-\mu (s-t_i)}\xi_2(t_i,i)\\
	&+ \int_{t_i}^{s}  e^{-\mu (s-\tilde{s})} \max (0,-x(\tilde{s},i)) d\tilde{s}. 
	\end{aligned}
	\label{eq:thresholdTime_2}
\end{equation} 
Consequently, when $x(t_i,i) >0$, for all $s \in [t_i, t_{i+1}]$, 
\begin{equation}
	\begin{aligned}
		&\xi_1(s,i) = e^{-\mu (s-t_i)}\xi_1(t_i,i)+ \int_{t_i}^{s} e^{-\mu (s-\tilde{s})} x(\tilde{s},i) d\tilde{s},\\
		&\xi_2(s,i) = e^{-\mu (s-t_i)}\xi_2(t_i,i). 
	\end{aligned}
	\label{eq:thresholdTime_pos}
\end{equation}
Conversely, when $x(t_i,i) <0$, for  all $s \in [t_i, t_{i+1}]$, we have
\begin{equation}
	\begin{aligned}
		&\xi_1(s,i) = e^{-\mu (s-t_i)}\xi_1(t_i,i),\\
		&\xi_2(s,i) = e^{-\mu (s-t_i)}\xi_2(t_i,i)- \int_{t_i}^{s} e^{-\mu (s-\tilde{s})} x(\tilde{s},i) d\tilde{s},
	\end{aligned}
	\label{eq:thresholdTime_neg}
\end{equation}
and if $x(t_i,i) = 0$, then for  all $s \in [t_i, t_{i+1}]$,
\begin{equation}
	\begin{aligned}
		&\xi_1(s,i) = e^{-\mu (s-t_i)}\xi_1(t_i,i), 
		\
		\xi_2(s,i) = e^{-\mu (s-t_i)}\xi_2(t_i,i).
	\end{aligned}
	\label{eq:thresholdTime_0}
\end{equation}

Note that, for each $ i \in \{0,1,\dots,j\}$, $q(t_{i+1}, i) \in \mathcal{D} = \mathcal{D}_1 \cup \mathcal{D}_2$. Let $q(t_{i+1},i) \in \mathcal{D}_\ell$, with $\ell \in \{1,2\}$, then, from \eqref{eq:JumpMap_1}, \eqref{eq:JumpMap_2} we have $\xi_\ell(t_{i+1},i+1)=0$ and $\xi_{3-\ell}(t_{i+1},i+1)=\xi_{3-\ell}(t_{i+1},i)$. Note that, since $n_x = 1$, from \eqref{eq:StateFlow_proofTh1} we have $\sign{x(t_{i+1},i)} = \sign{x(t_i,i)}$ and, thus, from \eqref{eq:jumpSet_i}, when $x(t_i,i) >0$, $q(t_{i+1}, i) \in \mathcal{D}_1$ and $\ell = 1$. In contrast, when $x(t_i,i) <0$, $q(t_{i+1}, i) \in \mathcal{D}_2$ and $\ell = 2$. Consequently,
%
 from \eqref{eq:thresholdTime_pos}-\eqref{eq:thresholdTime_0} we have $\xi_{3-\ell}(t_{i+1},i) = e^{-\mu(t_{i+1} - t_i)}\xi_{3-\ell}(t_{i},i)$. 
Moreover, 
since $\xi_\ell(0,0) = \xi_{3-\ell}(0,0) = 0$, and using \eqref{eq:jumpMap}-\eqref{eq:JumpMap_2}, 
we have that, for each $i \in \{0,1, \dots, j\}$, $\xi_\ell(t_i,i) = \xi_{3-\ell}(t_i,i) = 0$, and
for  all $s \in [t_i, t_{i+1}]$, $\xi_{3-\ell}(s, i) = 0$. Consequently, 
\eqref{eq:thresholdTime_pos}-\eqref{eq:thresholdTime_0} become
\begin{equation}
	\begin{aligned}
		&\xi_1(s,i) = \int_{t_i}^{s}e^{-\mu (s-\tilde{s})} x(\tilde{s},i) d\tilde{s}, 
		\ \xi_2(s,i) = 0 
	\end{aligned}
	\label{eq:thresholdTime_pos_bis}
\end{equation}
when $x(t_i,i) >0$, 
\begin{equation}
	\begin{aligned}
		&\xi_1(s,i) = 0, 
		\ \xi_2(s,i) = - \int_{t_i}^{s}e^{-\mu (s-\tilde{s})} x(\tilde{s},i) d\tilde{s},
	\end{aligned}
	\label{eq:thresholdTime_neg_bis}
\end{equation}
when  $x(t_i,i) <0$ and 
\begin{equation}
	\begin{aligned}
		&\xi_1(s,i) =
		\xi_2(s,i) = 0,
	\end{aligned}
	\label{eq:thresholdTime_0_bis}
\end{equation}
when $x(t_i,i) = 0$. The last equation, together with \eqref{eq:jumpSet}-\eqref{eq:jumpSet_i},  implies that, when $x(t_i,i) = 0$, $\sup_j \dom q = i$ and, from  \eqref{eq:StateFlow_proofTh1}, $x(t,j) = 0$ for all $(t,j) \in \dom q$, with $t \geq t_i$. For the remaining of the proof we consider the case  $x(t_i,i) \neq 0$.

Define $\xi:= \max(\xi_1, \xi_2)$, then, from \eqref{eq:StateFlow_proofTh1} and \eqref{eq:thresholdTime_pos_bis}-\eqref{eq:thresholdTime_0_bis}, 
\begin{equation}
	\xi(s,i) =  \int_{t_i}^{s} e^{-\mu (s-\tilde{s})} e^{a(\tilde{s}-t_i)}|x(t_i,i)| d\tilde{s}. 
	\label{eq:thresholdTime2}
\end{equation} 
In addition, \eqref{eq:jumpSet}-\eqref{eq:jumpSet_i} imply, when $x(t_i,i) \neq 0$, 
\begin{equation}
	t_{i+1}:= \inf\{s \geq t_i: \xi(s,i) = \Delta\}. 
	\label{eq:thresholdTime3}
\end{equation}
Thus,  
\begin{equation}
	\begin{aligned}
		\xi(t_{i+1},i) &=  |x(t_i,i)|\int_{t_i}^{t_{i+1}} e^{-\mu (t_{i+1}-\tilde{s})} e^{a(\tilde{s}-t_i)}  d\tilde{s}\\
		&=  |x(t_i,i)|e^{-\mu t_{i+1} - at_i}\int_{t_i}^{t_{i+1}} e^{(\mu +a) \tilde{s}} d\tilde{s}\\
		& = \frac{|x(t_i,i)|}{\mu + a} e^{-\mu t_{i+1} - a t_i} (e^{(\mu+a)t_{i+1}} -e^{(\mu+a)t_i})\\
		& =  \frac{|x(t_i,i)|}{\mu + a} (e^{a(t_{i+1}- t_i)} - e^{-\mu (t_{i+1} - t_i)})\\
		& = \Delta.   
	\end{aligned}
	\label{eq:thresholdTime4}
\end{equation} 
%
Since $\mu \in \R_{\geq 0}$ and $t_{i+1} \geq t_i$, $e^{-\mu (t_{i+1} - t_i)} \in (0,1]$. Consequently, from \eqref{eq:thresholdTime4}, 
\begin{equation}
	\begin{aligned}
		\frac{|x(t_i,i)|}{\mu + a} (e^{a(t_{i+1}- t_i)} - 1) \leq \Delta, 
	\end{aligned}
\end{equation}
which implies 
\begin{equation}
	t_{i+1}-t_i \leq \frac{1}{a} \ln\left( \frac{\Delta (\mu + a)}{|x(t_i,i)|} +1\right). 
	\label{eq:thresholdTime5}
\end{equation}

On the other hand, for each $i \in \{0,1,\dots,j\}$, $q(t_{i+1},i) \in \mathcal{D}$, and from \eqref{eq:EqProofTh1_1}-\eqref{eq:EqProofTh1_2} we have
\begin{equation}
	\begin{aligned}
		&V(x(t_{i+1}, i+1)) = (x(t_{i+1}, i)- \alpha)^2 \;\text{when} \ q(t_{i+1},i) \in \mathcal{D}_1\\
		&V(x(t_{i+1}, i+1)) = (x(t_{i+1}, i)+ \alpha)^2 \;  \text{when} \ q(t_{i+1},i) \in \mathcal{D}_2,\\
	\end{aligned}
	\label{eq:EqProofTh1_time}
\end{equation}
which are equivalent to 
\begin{equation}
	V(x(t_{i+1}, i+1)) = (x(t_{i+1}, i)- \alpha \sign{x(t_{i+1}, i)})^2. 
	\label{eq:EqProofTh1_2time}
\end{equation}


%
We first consider the case where $|x(t_i, i)| \in (\alpha, \sigma\Psi)$, $i \in \{0, 1,\dots, j\}$. 
From~\eqref{eq:StateFlow_proofTh1} we have
\begin{equation}
	|x(t_{i+1}, i)| = |e^{a(t_{i+1}-t_i)}x(t_i,i)| \geq |x(t_i,i)| > \alpha. 
\end{equation}
%
Consequently, from \eqref{eq:EqProofTh1_2time} and since $\alpha >0$ and $\sigma \in \left[\frac{(\rho + 1)^2 -1}{(\rho + 1)^2},1\right)$,  for each $i \in \{0,1,\dots,j\}$ such that $|x(t_i, i)| \in (\alpha, \sigma\Psi)$,  
\begin{equation}
	\begin{aligned}
		V(x(t_{i+1},i+1)) 
		&= (e^{a(t_{i+1}-t_i)}x(t_i,i)-\alpha \sign{x(t_i,i)})^2\\
		&\leq \frac{e^{a(t_{i+1}-t_i)}\sigma \Psi-\alpha}{e^{a(t_{i+1}-t_i)}\sigma \Psi} x^2(t_{i+1},i)\\
		&\leq \frac{e^{a(t_{i+1}-t_i)}\sigma \Psi-\alpha}{e^{a(t_{i+1}-t_i)} \Psi} V(x(t_{i+1}, i)).
	\end{aligned}
	\label{eq:EqProofTh1_4} 
\end{equation}
By merging \eqref{eq:EqProofTh1_3} and \eqref{eq:EqProofTh1_4} 
we have that for each $i \in \{0,1,\dots,j\}$ such that  $|x(t_i, i)| \in (\alpha, \sigma\Psi)$,  
\begin{equation}
	\begin{aligned}
		&V(x(t_{i+1},i+1))\\
		& \quad \leq \frac{e^{a(t_{i+1}-t_i)}\sigma \Psi-\alpha}{e^{a(t_{i+1}-t_i)} \Psi} e^{2a(t_{i+1}-t_i)}V(x(t_i,i))\\
		& \quad = \frac{e^{a(t_{i+1}-t_i)}\sigma \Psi-\alpha}{\Psi} e^{a(t_{i+1}-t_i)}V(x(t_i,i)).\\ 
	\end{aligned}
	\label{eq:EqProofTh1_5} 
\end{equation}
By substituting \eqref{eq:thresholdTime5} in the last inequality, we obtain
\begin{equation}
	\begin{aligned}
		&V(x(t_{i+1},i+1)) \\
		&\quad \leq \frac{\left( \frac{\Delta (\mu+a)}{|x(t_i,i)|} +1\right)\sigma \Psi-\alpha}{\Psi} \left( \frac{\Delta (\mu+a)}{|x(t_i,i)|} +1\right)V(x(t_i,i)).\\
	\end{aligned}
	\label{eq:EqProofTh1_6} 
\end{equation}
Using   $\Delta \in \left(0,\frac{ \rho\alpha}{\mu +a}\right]$, $|x(t_i, i)| \in (\alpha, \sigma\Psi)$ and $ \Psi = \frac{\rho +1}{(\rho+1)^2-1}\alpha$, we have, for each $i \in \{0, 1, \dots, j\}$, 
\begin{equation}
	\begin{aligned}
		&V(x(t_{i+1},i+1)) \\
		&\quad \leq \frac{(\frac{\rho\alpha}{|x(t_i,i)|} +1)\sigma \Psi-\alpha}{\Psi} \left(\frac{\rho\alpha}{|x(t_i,i)|} +1\right)V(x(t_i,i)).\\
		&\quad \leq \frac{(\rho +1)\sigma \Psi-\alpha}{\Psi} \left(\rho +1\right)V(x(t_i,i))\\
		&\quad =(1 - (1-\sigma)\left(\rho +1\right)^2)V(x(t_i,i))\\
		&\quad =\tilde{\gamma} V(x(t_i,i)),
	\end{aligned}
	\label{eq:EqProofTh1_7} 
\end{equation}
with $\tilde{\gamma}:=  (1 - (1-\sigma)\left(\rho +1\right)^2) \in (0,1)$. 
From \eqref{eq:EqProofTh1_7} and using $V(x) = x^2$ we have that for each $i \in \{0,1, \dots, j\}$ such that $|x(t_i, i)| \in (\alpha, \sigma\Psi)$, 
	$x(t_{i+1}, i+1)^2 \leq \tilde\gamma x(t_i, i)^2$. 
Consequently, 
\begin{equation}
	|x(t_{i+1}, i+1)| \leq \gamma |x(t_i, i)|,
	\label{eq:EqProofTh1_itemIII_1}
\end{equation}
where $\gamma=  \sqrt{\tilde\gamma} \in (0,1)$. 
%
In addition, 
from \eqref{eq:jumpMap}-\eqref{eq:JumpMap_2}
we have that, for all $s \in [t_{i}, t_{i+1}]$ such that $|x(t_i, i)| \in (\alpha, \sigma\Psi)$,  
\begin{equation}
	\begin{aligned}
		|x(s, i)| 
		&\leq \gamma |x(t_i, i)| + \alpha.\\
	\end{aligned}
	\label{eq:EqProofTh1_itemIII_2}
\end{equation}

We now consider the case when $|x(t_i, i)| \leq \alpha$, $i \in \{0,1,\dots,j\}$. 
We need to distinguish two cases. First consider the case where there exists $t \in [t_i, t_{i+1}]$ such that $|x(t, i)|> \alpha$.  Define $\tilde{t}:= \inf\{t\geq t_i:|x(t, i)|= \alpha\}$. 
Following similar steps as in the case where $|x(t, i)|\in (\alpha, \sigma \Psi)$ we obtain
\begin{equation}
	\begin{aligned}
		V(x(t_{i+1},i+1)) 
  \leq \tilde{\gamma} V(x(\tilde{t},i))
  = \tilde\gamma x(\tilde{t},i)^2
  = \tilde\gamma \alpha^2
  \leq \alpha^2.
	\end{aligned}
	\label{eq:eqProofTh1Item2_1}
\end{equation}
Using $V(x) = x^2$ we have 
	$x(t_{i+1}, i+1)^2 \leq \alpha^2$,
which implies
	$|x(t_{i+1}, i+1)| \leq \alpha$.
Consequently, following similar steps as in the case where  $|x(t_i, i)| \in (\alpha, \sigma\Psi)$, from \eqref{eq:jumpMap}-\eqref{eq:JumpMap_2} and \eqref{eq:eqProofTh1Item2_1}, 
 we have, for all $s \in [\tilde{t}, t_{i+1}]$,
	$|x(s,i)| \leq 2\alpha$.
Moreover, for all $s \in [t_i, \tilde{t}]$, from the definition of $\tilde{t}$, we have
	$|x(s,i)| \leq \alpha$.  
Consequently, from the last two inequalities, we have that 
\begin{equation}
	|x(s,i)| \leq 2\alpha
	\label{eq:StateFlow_itemIII_4}
\end{equation}
holds for all $s \in [t_i, t_{i+1}]$. 

On the other hand, when $\tilde{t}$ does not exists, 
we have that for all $s \in [t_i, t_{i+1}]$,
\begin{equation}
	|x(s, i)|\leq \alpha.
	\label{eq:StateFlow_itemIII_5}
\end{equation} 

Merging \eqref{eq:StateFlow_itemIII_4} and \eqref{eq:StateFlow_itemIII_5} we have that for each $i \in \{0,1, \dots, j\}$ such that $|x(t_i,i)| \leq \alpha$, for all $s \in [t_i, t_{i+1}]$, 
\begin{equation}
	|x(s,i)| \leq 2\alpha.
	\label{eq:StateFlow_itemIII_6}
\end{equation}

Therefore, from \eqref{eq:EqProofTh1_itemIII_2} and \eqref{eq:StateFlow_itemIII_6} we have that for all $(t,j) \in \dom q$ 
\begin{equation}
	|x(t,j)| \leq \gamma^j |x(0, 0)| + 2\alpha, 
	\label{eq:ProofTh1_itemIII_final}
\end{equation}
which concludes the proof of item \textit{(i)}. 

We now prove item \textit{(ii)}. 
From \eqref{eq:jumpSet}-\eqref{eq:jumpSet_i} we have, for each $i \in \{0,1,\dots,j\}$ and  all $s \in (t_i, t_{i+1})$, 
\begin{equation}
	t_{i+1}= \inf\{s \geq t_i: \xi_1(s,i) = \Delta \text{ or } \xi_2(s,i) = \Delta\}. 
	\label{eq:thresholdtime_DwellTimeProof}
\end{equation}

From \eqref{eq:thresholdFlow} we have,  for each $i \in \{0,1,\dots,j\}$ and  all $s \in (t_i, t_{i+1})$, 
\begin{equation}
	\begin{aligned}
		&\frac{d}{ds} \xi_1(s,i) = -\mu \xi_1(s,i) +  \max(0,x(s,i)), \\ 
		&\frac{d}{ds} \xi_2(s,i) = -\mu \xi_2(s,i) +  \max(0,-x(s,i)).  
	\end{aligned}
	\label{eq:dwellTimeProofDerivative}
\end{equation}
From \eqref{eq:thresholdTime_pos_bis}-\eqref{eq:thresholdTime_0_bis} we have that,  for each $i \in \{0,1,\dots,j\}$ and  all $s \in (t_i, t_{i+1})$, 
\begin{equation}
	\begin{aligned}
		&\xi_2(s,i) = 0 \quad \text{when }  x(t_i,i)>0\\
		&\xi_1(s,i) = 0 \quad \text{when }  x(t_i,i)<0\\
		&\xi_1(s,i) = \xi_2(s,i) = 0 \quad \text{when }  x(t_i,i)=0.\\
	\end{aligned}
	\label{eq:dwellTimeProofDerivative2}
\end{equation}
Consequently, for each $i \in \{0,1,\dots,j\}$ and  all $s\in(t_i, t_{i+1})$,
\begin{equation}
	\begin{aligned}
		&\frac{d}{ds} \xi_2(s,i) = 0 \quad \text{when }  x(t_i,i)>0\\
		&\frac{d}{ds} \xi_1(s,i) = 0 \quad \text{when }  x(t_i,i)<0\\
		&\frac{d}{ds} \xi_1(s,i) =\frac{d}{ds} \xi_2(s,i) = 0 \quad \text{when }  x(t_i,i)=0.\\
	\end{aligned}
	\label{eq:dwellTimeProofDerivative3}
\end{equation}
Therefore, from \eqref{eq:dwellTimeProofDerivative}-\eqref{eq:dwellTimeProofDerivative3} 
we have, for each $i \in \{0,1,\dots,j\}$ and  all $s \in (t_i, t_{i+1})$, 
\begin{equation}
	\begin{aligned}
		\frac{d}{ds} \xi(s,i) &= -\mu \xi(s,i) +  |x(s,i)|,\\
	\end{aligned}
	\label{eq:dwellTimeProofDerivative4}
\end{equation}
where $\xi = \max(\xi_1, \xi_2)$. Moreover, from \eqref{eq:ProofTh1_itemIII_final} we have that, for each $i \in \{0,1,\dots,j\}$ and  all $s \in (t_i, t_{i+1})$, 
\begin{equation}
	\begin{aligned}
		|x(s,i)|&\leq \gamma^i |x(0, 0)| + 2\alpha 
		\leq \Psi + 2\alpha 
		= \Upsilon. 
	\end{aligned}
	\label{eq:boundedState}
\end{equation}
From \eqref{eq:dwellTimeProofDerivative4}-\eqref{eq:boundedState}, since $\mu, \xi \in \R_{> 0}$, we have, for each $i \in \{0,1,\dots,j\}$ and  all $s \in (t_i, t_{i+1})$, 
		$\displaystyle \frac{d}{ds} \xi(s,i) \leq |x(s,i)| 
		 \leq \Upsilon.$  
Integrating the last inequality and applying the comparison principle \cite[Lemma 3.4]{khalil2002nonlinear}, we obtain, for all $s \in (t_i, t_{i+1})$, 
\begin{equation}
	\begin{aligned}
		\xi(s,i) &\leq \int_{t_i}^{s} \Upsilon d\tilde{s} 
		= \Upsilon(s-t_i). 
	\end{aligned}
	\label{eq:dwelTimeProofXi}
\end{equation}
Moreover, in view of \eqref{eq:dwellTimeProofDerivative2}-\eqref{eq:dwellTimeProofDerivative3}, \eqref{eq:thresholdtime_DwellTimeProof} is equivalent to 
\begin{equation}
	t_{i+1}= \inf\{s \geq t_i: \xi(s,i) = \Delta\}. 
	\label{eq:thresholdtime_DwellTimeProof2}
\end{equation}
Thus, from \eqref{eq:dwelTimeProofXi} and \eqref{eq:thresholdtime_DwellTimeProof2} we obtain, for each $i \in \{0,1, \dots, j\}$, 
	 $t_{i+1} - t_i \geq \frac{\Delta}{\Upsilon} 
		= \tau, $
which implies that the solution $q$ has a dwell-time $\tau \in \R_{> 0}$. 
This concludes the proof. 
%
%
%
\hfill$\blacksquare$

\bibliography{IEEEabrv,bibliography}

\begin{thebibliography}{10}

\bibitem{mead1990neuromorphic}
C.~A. Mead, ``Neuromorphic electronic systems,'' {\em Proceedings of the IEEE},
  vol.~78, no.~10, pp.~1629--1636, 1990.

\bibitem{gallego2022event}
G.~Gallego, T.~Delbrück, G.~Orchard, C.~Bartolozzi, B.~Taba, A.~Censi,
  S.~Leutenegger, A.~J. Davison, J.~Conradt, K.~Daniilidis, and D.~Scaramuzza,
  ``Event-based vision: A survey,'' {\em IEEE Transactions on Pattern Analysis
  and Machine Intelligence}, vol.~44, no.~1, pp.~154--180, 2022.

\bibitem{van2018organic}
Y.~van De~Burgt, A.~Melianas, S.~T. Keene, G.~Malliaras, and A.~Salleo,
  ``Organic electronics for neuromorphic computing,'' {\em Nature Electronics},
  vol.~1, no.~7, pp.~386--397, 2018.

\bibitem{krauhausen2021organic}
I.~Krauhausen, D.~A. Koutsouras, A.~Melianas, S.~T. Keene, K.~Lieberth,
  H.~Ledanseur, R.~Sheelamanthula, A.~Giovannitti, F.~Torricelli, I.~Mcculloch,
  P.~W.~M. Blom, A.~Salleo, Y.~van~de Burgt, and P.~Gkoupidenis, ``Organic
  neuromorphic electronics for sensorimotor integration and learning in
  robotics,'' {\em Science Advances}, vol.~7, no.~50, p.~eabl5068, 2021.

\bibitem{shrestha2022survey}
A.~Shrestha, H.~Fang, Z.~Mei, D.~P. Rider, Q.~Wu, and Q.~Qiu, ``A survey on
  neuromorphic computing: Models and hardware,'' {\em IEEE Circuits and Systems
  Magazine}, vol.~22, no.~2, pp.~6--35, 2022.

\bibitem{schuman2017survey}
C.~D. Schuman, T.~E. Potok, R.~M. Patton, J.~D. Birdwell, M.~E. Dean, G.~S.
  Rose, and J.~S. Plank, ``A survey of neuromorphic computing and neural
  networks in hardware,'' {\em arXiv preprint arXiv:1705.06963}, 2017.

\bibitem{izhikevich2004model}
E.~M. Izhikevich, ``Which model to use for cortical spiking neurons?,'' {\em
  IEEE Transaction on Neural Networks}, vol.~15, no.~5, pp.~1063--1070, 2004.

\bibitem{izhikevich2007dynamical}
E.~M. Izhikevich, {\em Dynamical systems in neuroscience}.
\newblock MIT press, 2007.

\bibitem{drion2015neuronal}
G.~Drion, T.~O'Leary, J.~Dethier, A.~Franci, and R.~Sepulchre, ``Neuronal
  behaviors: A control perspective,'' {\em IEEE Conference on Decision and
  Control, \textnormal{Osaka, Japan}}, pp.~1923--1944, 2015.

\bibitem{hodgkin1952quantitative}
A.~L. Hodgkin and A.~F. Huxley, ``A quantitative description of membrane
  current and its application to conduction and excitation in nerve,'' {\em The
  Journal of physiology}, vol.~117, no.~4, p.~500, 1952.

\bibitem{fitzhugh1961impulses}
R.~FitzHugh, ``Impulses and physiological states in theoretical models of nerve
  membrane,'' {\em Biophysical journal}, vol.~1, no.~6, pp.~445--466, 1961.

\bibitem{hindmarsh1984model}
J.~L. Hindmarsh and R.~Rose, ``A model of neuronal bursting using three coupled
  first order differential equations,'' {\em Proceedings of the Royal society
  of London. Series B. Biological sciences}, vol.~221, no.~1222, pp.~87--102,
  1984.

\bibitem{izhikevich2010hybrid}
E.~M. Izhikevich, ``Hybrid spiking models,'' {\em Philosophical Transactions of
  the Royal Society A: Mathematical, Physical and Engineering Sciences},
  vol.~368, no.~1930, pp.~5061--5070, 2010.

\bibitem{lapicque1907recherches}
L.~Lapicque, ``Recherches quantitatives sur l’excitation \'electrique des
  nerfs trait\'ee comme une polarization,'' {\em J. Physiol. Pathol. Gen.},
  vol.~9, pp.~620--635, 1907.

\bibitem{abbott1999lapicque}
L.~F. Abbott, ``Lapicque’s introduction of the integrate-and-fire model
  neuron (1907),'' {\em Brain research bulletin}, vol.~50, no.~5-6,
  pp.~303--304, 1999.

\bibitem{gerstner2002spiking}
W.~Gerstner and W.~M. Kistler, {\em Spiking neuron models: Single neurons,
  populations, plasticity}.
\newblock Cambridge University Press, 2002.

\bibitem{heemels2010hybrid}
W.~P. M.~H. Heemels, B.~De~Schutter, J.~Lunze, and M.~Lazar, ``Stability
  analysis and controller synthesis for hybrid dynamical systems,'' {\em
  Philosophical Transactions of the Royal Society A: Mathematical, Physical and
  Engineering Sciences}, vol.~368, no.~1930, pp.~4937--4960, 2010.

\bibitem{goebel2012hybrid}
R.~Goebel, R.~G. Sanfelice, and A.~R. Teel, {\em Hybrid Dynamical Systems:
  Modeling, Stability, and Robustness}.
\newblock New Jersey, USA, Princeton University Press, 2012.

\bibitem{deweerth1990neuron}
S.~P. DeWeerth, L.~Nielsen, C.~A. Mead, and K.~J. {\AA}str{\"o}m, ``A
  neuron-based pulse servo for motion control,'' in {\em IEEE International
  Conference on Robotics and Automation}, pp.~1698--1703, 1990.

\bibitem{deweerth1991simple}
S.~P. DeWeerth, L.~Nielsen, C.~A. Mead, and K.~J. {\AA}str{\"o}m, ``A simple
  neuron servo,'' {\em IEEE Transactions on Neural Networks}, vol.~2, no.~2,
  pp.~248--251, 1991.

\bibitem{ribar2021neuromorphic}
L.~Ribar and R.~Sepulchre, ``Neuromorphic control: Designing multiscale
  mixed-feedback systems,'' {\em IEEE Control Systems Magazine}, vol.~41,
  no.~6, pp.~34--63, 2021.

\bibitem{sepulchre2022spiking}
R.~Sepulchre, ``Spiking control systems,'' {\em Proceedings of the IEEE},
  vol.~110, no.~5, pp.~577--589, 2022.

\bibitem{singh2016stabilization}
P.~Singh, S.~Z. Yong, J.~Gregoire, A.~Censi, and E.~Frazzoli, ``Stabilization
  of linear continuous-time systems using neuromorphic vision sensors,'' {\em
  IEEE Conference on Decision and Control, \textnormal{Las Vegas, USA}},
  pp.~3030--3036, 2016.

\bibitem{singh2018regulation}
P.~Singh, S.~Z. Yong, and E.~Frazzoli, ``Regulation of linear systems using
  event-based detection sensors,'' {\em IEEE Transactions on Automatic
  Control}, vol.~64, no.~1, pp.~373--380, 2018.

\bibitem{aastrom1999comparison}
K.~J. {\AA}str{\"o}m and B.~Bernhardsson, ``Comparison of periodic and event
  based sampling for first-order stochastic systems,'' {\em IFAC Proceedings
  Volumes}, vol.~32, no.~2, pp.~5006--5011, 1999.

\bibitem{tabuada2007event}
P.~Tabuada, ``Event-triggered real-time scheduling of stabilizing control
  tasks,'' {\em IEEE Transaction on Automatic Control}, vol.~52, no.~9,
  pp.~1680--1685, 2007.

\bibitem{heemels2012introduction}
W.~P. M.~H. Heemels, K.~H. Johansson, and P.~Tabuada, ``An introduction to
  event-triggered and self-triggered control,'' {\em IEEE Conference on
  Decision and Control, \textnormal{Maui, USA}}, pp.~3270--3285, 2012.

\bibitem{postoyan2014framework}
R.~Postoyan, P.~Tabuada, D.~Ne{\v{s}}i{\'c}, and A.~Anta, ``A framework for the
  event-triggered stabilization of nonlinear systems,'' {\em IEEE Transactions
  on Automatic Control}, vol.~60, no.~4, pp.~982--996, 2014.

\bibitem{aaarzen1999simple}
K.-E. {\AA}rz{\'e}n, ``A simple event-based {PID} controller,'' {\em IFAC
  Proceedings Volumes}, vol.~32, no.~2, pp.~8687--8692, 1999.

\bibitem{brockett2000quantized}
R.~W. Brockett and D.~Liberzon, ``Quantized feedback stabilization of linear
  systems,'' {\em IEEE transactions on Automatic Control}, vol.~45, no.~7,
  pp.~1279--1289, 2000.

\bibitem{liberzon2003hybrid}
D.~Liberzon, ``Hybrid feedback stabilization of systems with quantized
  signals,'' {\em Automatica}, vol.~39, no.~9, pp.~1543--1554, 2003.

\bibitem{nesic2009unified}
D.~Nesic and D.~Liberzon, ``A unified framework for design and analysis of
  networked and quantized control systems,'' {\em IEEE Transactions on
  Automatic control}, vol.~54, no.~4, pp.~732--747, 2009.

\bibitem{de2009robust}
C.~De~Persis, ``Robust stabilization of nonlinear systems by quantized and
  ternary control,'' {\em Systems \& Control Letters}, vol.~58, no.~8,
  pp.~602--608, 2009.

\bibitem{khalil2002nonlinear}
H.~K. Khalil, {\em Nonlinear Systems}, vol.~3.
\newblock Prentice hall Upper Saddle River, NJ, 2002.

\end{thebibliography}

\end{document}